\documentclass[twocolumn,showpacs,preprintnumbers,amsmath,amssymb]{revtex4}

\usepackage{graphicx}
\usepackage{dcolumn}
\usepackage{bm}
\usepackage{epsfig}
\newcommand{\be}{\begin{eqnarray}}
\newcommand{\ee}{\end{eqnarray}}

\begin{document}
\title{One-dimensional multicomponent fermions with delta function
    interaction in strong and weak coupling limits: Two-component Fermi
    gas}

\author{ Xi-Wen Guan}
\affiliation{Department of
Theoretical Physics, Research School of Physics and Engineering,
Australian National University, Canberra ACT 0200, Australia}
\author{ Zhong-Qi Ma}
\affiliation{Institute of High Energy Physics, Chinese Academy of Sciences, Beijing 100049, China}

\date{\today}

\begin{abstract}
 The Fredholm equations  for one-dimensional  two-component Fermions with repulsive and with
attractive  delta-function interactions  are  solved by an asymptotic expansion  for A) strong repulsion, B) weak repulsion, C) weak attraction and D) strong attraction.  Consequently, we obtain the first few terms of the expansion  of ground state energy for the Fermi gas with  polarization  for  these regimes.  
We also prove that  the two sets of the Fredhom equations for weakly repulsive and attractive interactions are identical  as long as the integration boundaries match each other between the two sides. Thus the asymptotic expansions of the energies of the repulsive and attractive Fermions  are identical to all orders   in  this region. The identity of the asymptotic expansions  may not mean that the energy analytically connects. 
 \end{abstract}

\maketitle

\section{Introduction} 

One-dimensional (1D)  Fermi  gases with delta-function interaction  are  important exactly solvable quantum many-body systems and have had tremendous impact in quantum statistical mechanics.  The two-component  delta-function interaction Fermi gas with arbitrary polarization  was exactly solved by Yang \cite{Yang} using the Bethe ansatz hypothesis in 1967.   Sutherland \cite{Sutherland} generalized the ansatz to solve the 1D multicomponent Fermi gas with delta-function interaction  in 1968.  The study of multicomponent attractive Fermi gases  was initiated   by  Yang \cite{Yang-a} and by Takahashi \cite{Takahashi-a} in 1970.  Since then exactly solvable models have been extensively studied by a variety of methods developed in the context of mathematical physics, see \cite{Yang-a,Hubbard,Takahashi-b}. In particular, recent breakthrough experiments on trapped  fermionic atoms confined to one dimension \cite{Liao}  have provided a better understanding of significant quantum statistic effects and novel pairing nature  in quantum many-body systems. The observed result is  seen to be in good  agreement with the result obtained using the  analysis of exactly solved models \cite{Orso,HuiHu,Guan,Wadati,Mueller}.

Despite the  Bethe ansatz equations  for  the 1D two-component  delta-function interaction Fermi gas with arbitrary polarization  were found  long ago \cite{Yang}, it was not  until much later that this model began to receive more attention from cold atoms \cite{Zwerger}.
The asymptotic ground state energy  of  the   Fermi gas with  polarization was studied by  the discrete Bethe ansatz equations in a strongly and weakly  interacting regimes in \cite{BBGO}. But it turns out that the asymptotic expansion of the discrete Bethe ansatz equations can  only be controllable up to the leading order correction to the interaction energy. 
However, the fundamental physics of integrable models are usually determined by the set of  the generalised Fredholm integral equations in the thermodynamic limit, see an insightful article  by Yang \cite{Yang-insight}. The solutions of  the  Fredholm equations have not been thoroughly investigated analytically except few limiting cases \cite{Zwerger,Ma} and numerical result  \cite{Yang-Ma,Orso,HuiHu,colome}. It  is usual  a difficult  task to solve those Fredeholm equations analytically.  It is highly desirable to find a systematic way to treat the generalized Fredholm equations. 

In the present paper,  we develop   a systematic  method  to solve asymptotically the Fredholm equations for 1D two-component   Fermi gas  with delta function interaction and with  polarization in  four regimes:  A) strongly repulsive regime; B) weakly repulsive regime; C) weakly attractive regime and D) strongly attractive regime. The first few terms of the expansion  of the ground state energy for the Fermi gas with  polarization are obtained explicitly for   these regimes. We also address the analytical behaviour of the ground state energy  at vanishing interaction strength. This method which we develop  can be directly applied to 1D multicomponent Fermi gases. The result of 1D $\kappa$-component fermions will be reported in the second paper of this study \cite{note}.

\section{ The Fredholm equations}

The Hamiltonian for the 1D $N$-body  problem \cite{Yang,Gaudin} 
\begin{equation}
H=-\frac{\hbar^{2}}{2m}\sum_{i=1}^{N}\frac{\partial^{2}}{\partial
x_{i}^{2}}+g_{1D}\sum_{1\leq i<j\leq
N}\delta(x_{i}-x_{j}).\label{Ham}
\end{equation}
 describes $N$  Fermions of the same mass $m$ with two internal spin  states  confined to a 1D
system of length $L$ interacting via a $\delta$-function potential.
 For an irreducible representation 
$[2^{N_{\downarrow}},1^{N_{\uparrow}-N_{\downarrow}}]$, the Young tableau has two columns.
Where  $N_{\uparrow}$ and $N_{\downarrow}$  are the numbers of Fermions in  the two  hyperfine levels $|\uparrow \rangle $ and $|\downarrow \rangle$  such that $N=N_{\uparrow}+ N_{\downarrow}$. The
coupling constant $g_{1D}$ can be expressed in terms of the
interaction strength $c=-2/a_{1D}$ as $g_{1D}=\hbar^{2}c/m$ where
$a_{1D}$ is the effective 1D scattering length  \cite{Olshanii}. Let $2m=\hbar =1$ for our convenience. 
We define  a dimensionless interaction strength $\gamma = c / n$	for the physical analysis,	with	the linear density  $n = N / L$.  For repulsive
Fermions, $c>0$ and for attractive Fermions, $c<0$.

The energy eigenspectrum is given in terms of the quasimomenta $\left\{k_i\right\}$  of the fermions via
$E=\frac{\hbar ^2}{2m}\sum_{j=1}^Nk_j^2$,
which in terms of the function $e_b(x)=(x+\mathrm{i}{bc}/{2})/(x-\mathrm{i}{bc}/{2})$ 
satisfy the BA equations 
\begin{eqnarray}
&&\exp(ik_{i}L)=\prod_{\alpha=1}^{N_{\downarrow}}e_1\left(k_i-\lambda_{\alpha}\right),\nonumber\\
&&
\prod_{j=1}^{N}e_1\left(\lambda_{\alpha }-k_j \right)=-\prod_{\beta=1}^{N_{\downarrow}}e_2\left(\lambda_{\alpha}-\lambda_{\beta}  \right),
\label{BE-2}
\end{eqnarray}
where $i=1,2,\ldots, N$ and $\alpha=1,2,\ldots, N_{\downarrow}$. The parameters $\left\{\lambda_{\alpha}\right\}$ are the
rapidities for the internal hyperfine spin degrees of freedom. The fundamental physics of the model are determined by the set of transcendental equations  which can be transformed to the generalised Fredholm types of equations in the thermodynamic limit.  This transformation  was found by Yang and Yang in series of papers on the study of spin $XXZ$ model  in 1966, see an  insightful  article  \cite{Yang-insight}.

\subsection{Repulsive regime}

 For repulsive interaction,  it is  shown from (\ref{BE-2}) that the Bethe ansatz quasimomenta $\left\{ k_i\right\}$ are real, but all  $\left\{ \lambda_{\alpha }\right\}$ are real only for  the ground state, see Ref. \cite{Takahashi-b}. There are complex roots of $\lambda_{\alpha}$ called spin strings for excited states. In the thermodynamic limit, i.e., $L,N \to \infty$, $N/L$ is finite, the roots of the  Bethe ansatz equations (\ref{BE-2}) are dense enough in the parameter space.  Therefore we can define  the particle quasimomentum distribution function  $r_1(k_i)=1/[L(k_i-k_{i+1})]$ in the quasimomentum  space. Here $k_i$ and $k_{i+1}$ are two conjunction quaisimomentua.   Similarly,  the distribution function of the spin rapidity  is defined $r_2(\lambda_i)=1/[L(\lambda_i-\lambda_{i+1})]$ in spin parameter space.    In order to unify the notations in the Fredholm equations, we replace the parameter $\lambda$ by $k$ for the distribution function of the spin rapidity. 
 Thus the above Bethe ansatz equations  (\ref{BE-2})  can be written as the generalized Fredholm equations 
\begin{eqnarray}
{r_1}(k)&=&\frac{1}{2\pi}+ \int_{-B_2}^{B_2}K_1(k-k'){r_2}(k')dk', \label{BE2-r1}\\
{r_2}(k)&=&\int_{-B_1}^{B_1}K_1(k-k'){r_1}(k')dk\nonumber \\
&& - \int_{-B_2}^{B_2}K_2(k-k'){r_2}(k') dk'.\label{BE2-r2}
\end{eqnarray}
The associated   integration boundaries $B_1$, $B_2$   are determined by the relations 
\begin{eqnarray}
n:&\equiv &N/L=\int_{-B_1}^{B_1}{r_1}(k)dk,  \nonumber\\
 n_{\downarrow}:&\equiv& N_{\downarrow}/L=\int_{-B_2}^{B_2}{r_2}(k)dk,\label{repulsive-d}
\end{eqnarray} 
where  $n$ denotes  the linear density while $n_{\downarrow}$ is the density of spin-down Fermions. The boundary $B_1$ characterizes the Fermi point in the quasimomentum space whereas the boundary $B_2$ characterizes the spin rapidity distribution interval with respect to the  polarization.  They can be  obtained  by solving the equations in  $\ref{repulsive-d}$. 
In the above equations,  we denote the kernel  function as
\begin{equation}
K_{\ell }(x)=\frac{1}{2\pi}\frac{\ell c}{(\ell c/2)^2+x^2}. \label{a-r}
\end{equation}
The ground state energy per unit length is given by 
\begin{equation}
E=\int_{-B_1}^{B_1}k^2{r_1}(k) d k. \label{E2-r} 
\end{equation}
The magnetization per length  is defined by $s_z=(n-2n_{\downarrow})/2$. Through the boundary conditions (\ref{repulsive-d}),  the ground state energy (\ref{E2-r}) can be expressed as function of total particle density  $n$ and magnetization $s_z$. In the grand canonical ensemble,  we can also get the magnetic field $h$ and chemical potential $\mu$ via
\begin{equation}
h=2\frac{\partial E(n,s_z) }{\partial s_z}, \qquad \mu =\frac{\partial E(n,s_z)}{\partial n}.\label{chemical}
\end{equation}

\subsection{Attractive regime}

For attractive regime, i.e. $c<0$, it is found from (\ref{BE-2})  that complex string solutions of $k_i$ also satisfy the Bethe ansatz  equations. Thus the quasimomenta $\left\{ k_i\right\} $ of the fermions with different spins  form two-body bound states \cite{Takahashi-a,Gu-Yang},
i.e., $k_i=k_i'\pm   \mathrm{i} \frac{1}{2} c$, accompanied by the real spin parameter
 $k'_i$.  Here $i=1,\ldots,N_{\downarrow}$.  The excess fermions have real quasimomenta $\left\{ k_j\right\} $ with $j=1,\ldots, N-2N_{\downarrow}$.
Thus the Bethe ansatz equations  are transformed into the Fredholm equations regarding to  the densities of the pairs $\rho_2(k)$   and density of single Fermi atoms $\rho_1(k)$.    They  satisfy the following Fredholm equations \cite{Yang-a,Takahashi-a}
\begin{eqnarray}
\rho_1(k)&=&\frac{1}{2\pi}+\int_{-Q_2}^{Q_2}K_1(k-k' )\rho_2(k')dk'  \label{Fermi2-a1}\\
\rho_2(k)&=&\frac{2}{2\pi}+\int_{-Q_1}^{Q_1}K_1(k-k')\rho_1(k')dk'\nonumber\\
&&+\int_{-Q_2}^{Q_2}K_2(k-k')\rho_2(k')dk'. \label{Fermi2-a2}
\end{eqnarray}
Here $c<0$ in the kernel functions $K_\ell(x)$. Here the integration boundaries $Q_1$ and $Q_2$ are the Fermi points of the single particles and pairs, respectively. They are 
 determined  by
\begin{eqnarray}
n&\equiv:& \frac{N}{L}= 2\int_{-Q_2}^{Q_2}\rho_2(k)dk+\int^{Q_1}_{-Q_1}\rho_1(k)dk,\nonumber \\
n_{\downarrow}&\equiv: & \frac{N_{\downarrow}}{L}=\int_{-Q_2}^{Q_2}\rho_2(k)dk.\label{density-a}
\end{eqnarray}
The ground state energy per length is given by
\begin{equation}
E=\int_{-Q_2}^{Q_2}\left(2k^2-c^2/2\right)\rho_2(k)dk+\int_{-Q_1}^{Q_1}k^2\rho_1(k)dk. \label{Fermi2-E-a}
\end{equation}
In a similar way, the magnetic field and chemical potential can be determined from the relations (\ref{chemical}). 
In next section, we will discuss solutions and analytical behaviour of the Fredholm equations.

\section{Asymptotic solutions of  the Fredholm equations}

\subsection{Strong repulsion}
\begin{figure}[tbp]
\includegraphics[width=1.15\linewidth]{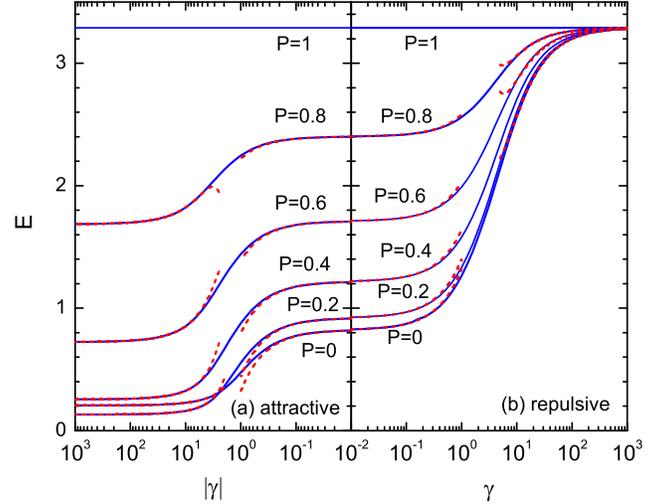}
\caption{ (Color online) The ground state energy per length  vs logarithmic $\gamma=cL/N$ in the unit of $ \frac{\hbar^2N^3}{2mL^2}$: comparison between the asymptotic solutions and numerical solutions of the Fredholm equations for polarization $P=0,\,0.2,\,0.4,\,0.6,\,0.8,\,1.0$.  In attractive regime, the biding energy $\varepsilon_{b} =-c^2/2$ was subtracted. The crossing of the two lowest curves  in attractive regimes indicates a  relation between  the critical polarization  and interaction, where the chemical potential of single Fermions excess the chemical potential of the pairs.    An excellent agreement between our asymptotic results and numerical plots is seen for  A) strong repulsion, B) weak repulsion, C) weak attraction and D) strong attraction}
\label{fig:Energy}
\end{figure}

\begin{figure}[tbp]
\includegraphics[width=1.00\linewidth]{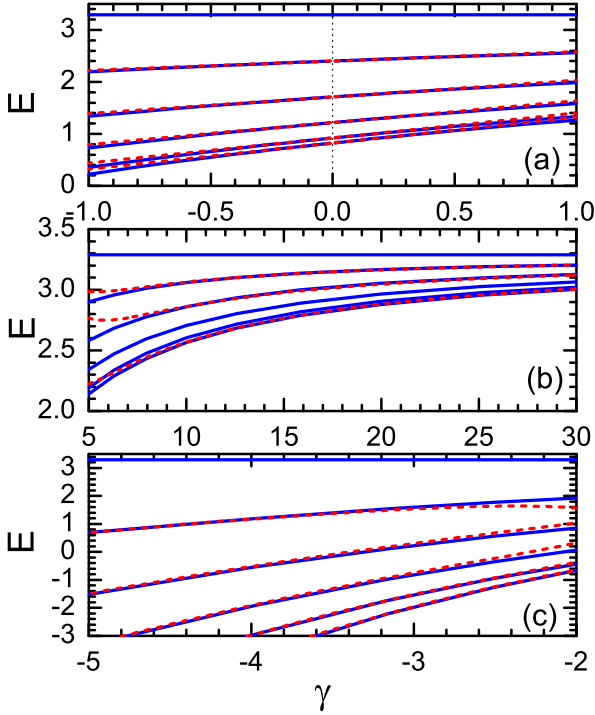}
\caption{(Color online) The ground state energy per length vs $\gamma=cL/N$ in the unit of $ \frac{\hbar^2N^3}{2mL^2}$: comparison between the asymptotic solutions and numerical solutions of the Fredholm equations for polarization $P=0,\,0.2,\,0.4,\,0.6,\,0.8,\,1.0$.   An excellent agreement between our asymptotic expansion  results and numerical plots is seen  in (a) for weakly repulsive and attractive regimes, (b) for strongly repulsive regime and (c) strongly attractive regime. 
 }
\label{fig:Energy2}
\end{figure}

The strong coupling condition   $cL/N\gg1$ naturally gives the condition $c\gg B_1$,  where the Fermi boundary $B_1\propto n\pi$ according to the Fermi statistics. For balanced case, the  numbers of spin-up and -down Fermions are equal. Thus there is no finite ``Fermi" points in spin parameter space, i.e.  the boundary $B_2\to \infty$.  Taking a  Taylor expansion with the kernel function $K_1(k-k')$  in  (\ref{BE2-r2}) at $k'=0$, we obtained with  an accuracy up to the order of $1/c^4$
\begin{eqnarray}
r_2(k)&\approx&nK_1(k)+\frac{E}{2\pi} \left[\frac{3c}{\left(\frac{c^2}{4}+k^2\right)^2} -\frac{c^3}{\left(\frac{c^2}{4}+k^2\right)^3}\right]\nonumber\\
&&-\int_{-\infty}^{\infty}K_2(k-k')r_2(k')dk'.\label{A2}
\end{eqnarray}
Here $E$ is the ground state energy per length. By taking Fourier transformation with (\ref{A2}), we may obtain  the distribution function 
\begin{equation}
\tilde{r}_2(\omega) =\left(n-\frac{E\omega^2}{2} \right)/\left( 2\cosh\frac{\omega c}{2}\right). \label{r2-e}
\end{equation}
Substituting  (\ref{r2-e}) into the Fredholm equation (\ref{BE2-r1}), we have 
\begin{eqnarray}
r_1(k)&=& \frac{1}{2\pi} +\frac{1}{2\pi}\int_{-\infty}^{\infty}e^{-\frac{1}{2}c|\omega|}\tilde{r}_2(\omega)e^{\mathrm{i}\omega k}d\omega\nonumber\\
&=&\frac{1}{2\pi}+\frac{n}{2\pi}\left( Y_0(k) -\frac{E}{2n}Y_2(k) \right)
\end{eqnarray}
where
\begin{equation}
Y_\alpha (k)\approx \int_{-\infty}^{\infty}\frac{e^{\mathrm{i}\omega k}\omega^{\alpha}d\omega}{1+e^{|\omega|c}}.\nonumber 
\end{equation}
After some algebra, we obtain 
\begin{equation}
Y_0(k)=\frac{2\ln 2}{c}-\frac{3k^2}{2c^3}\zeta(3), \qquad Y_2(k)\approx \frac{3}{c^3}\zeta(3).\nonumber
\end{equation}
Here $\zeta(z)$ is the Riemann's zeta function.  Then we obtain
\begin{equation}
r_1(k)=\frac{1}{2\pi}+\frac{n\ln 2}{\pi c}-\frac{3n\zeta(3)}{4\pi c^3}\left(k^2+\frac{E}{n} \right)+O(c^{-4}).\label{r-r}
\end{equation}
We see clearly that for strong repulsion  the distribution of $r_1(k)$ is very flat  and it is a constant up to the order of $1/c^3$ correction. This naturally suggests that the 1D Fermions with strong repulsion can be treated as an ideal particles with fractional statistics.  
Substituting (\ref{r-r}) into  linear density  (\ref{repulsive-d}) and energy (\ref{E2-r}), we obtain
\begin{eqnarray}
n&=&\frac{B_1}{\pi}\left(1+\frac{2n\ln 2}{c}-\frac{3E\zeta(3)}{2c^3}-\frac{n^3\pi^2\zeta(3)}{2c^3} \right),\nonumber \\
E&=&\frac{B_1^3}{3\pi}\left( 1+\frac{2n\ln 2}{c} -\frac{3E\zeta(3)}{2c^3}-\frac{9n^3\pi^2\zeta(3)}{10c^3} \right)\nonumber 
\end{eqnarray}
that give 
\begin{eqnarray}
B_1&=&n\pi\left[ 1-\frac{2\ln 2}{\gamma}\left(1-\frac{2\ln 2}{\gamma}\right) -\frac{8(\ln2)^3}{\gamma^3}\right.\nonumber\\
&&\left. +\frac{\pi^2\zeta(3)}{\gamma^3} \right]+O(c^{-4}),\\
E&=&\frac{n^3\pi^2}{3}\left[1- \frac{4\ln2}{\gamma} +\frac{12(\ln2)^2}{\gamma^2} -\frac{32(\ln2)^3}{\gamma^3}\right.\nonumber\\
&&\left. +\frac{8\pi^2\zeta(3)}{5\gamma^3} \right]+O(c^{-4}).\label{E-A-b}
\end{eqnarray}
We see that the energy is given in terms of the dimensionless strength $\gamma=c/n$. The leading order of $1/\gamma$ correction was found in \cite{Zwerger,BBGO}. Actually, the two sets of the Fredholm equations can be converted into dimensionless units. Therefore  the ground state energy can be written as  analytical functions of $\gamma$ except at  $\gamma = 0$.  
This ground state energy  is a  good approximation for the balanced Fermi gas with a  strongly repulsive  interaction (good agreement is seen for $cL/N>8$), see Figure \ref{fig:Energy} and Figure \ref{fig:Energy2}.  In these figures, solid lines are obtained from the ground state energy (\ref{E2-r}) and (\ref{Fermi2-E-a}) with 
the numerical solutions to the two sets of the Fredholm equations  (\ref{BE2-r1}) and    (\ref{BE2-r2})   for repulsive regime and    (\ref{Fermi2-a1}) and (\ref{Fermi2-a2})   for attractive regime. The dashed lines are plotted from the asymptotic ground state energy for  the four regimes. 

However, it seems to be extremely hard to obtain a close form of the ground state energy  of the gas with an arbitrary spin population imbalance in repulsive regime.  This is mainly because that the distribution function $r_2(k)$ spans in the region  $-B_2<k<B_2$,  where  the integration boundary $B_2$ can vary  from  zero to infinity as the polarization changes.  The integration boundary $B_2$  decreases  as the polarization increases. An intuitive way of understanding this point is that zero polarization corresponds to $B_2=\infty$ while fully-polarized case corresponds to $B_2=0$.  From dressed energy formalism \cite{Guan},   we can easily see this monotonic relation between the Fermi boundary and polarization  by analysing the band filling  under external field.  For high polarization and strong repulsion (i.e., $N_{\downarrow}\ll N$), we have the conditions  $c\gg B_1, \,B_2$, that  allows us to do the following Taylor expansion:
\begin{eqnarray}
r_2(k)&=&  \frac{1}{2\pi}\int_{-B_1}^{B_1} \frac{cr_1(k')}{\frac{c^2}{4}+k^2}\left[1-\frac{-2kk'+ k'^2}{\frac{c^2}{4}+k^2} +\cdots \right]dk'\nonumber\\
&& -\int_{-B_2}^{B_2}K_2(k-k') r_2(k')dk'\nonumber\\
&=&n\left(1-\frac{4E}{c^2n}\right)K_1(k)-n_{\downarrow}K_2(k)+O(c^{-4})\label{BE-Br21}.
\end{eqnarray}
Here we denote $n_{\downarrow}=N_{\downarrow}/L$.  We  notice that the leading order of the distribution function $r_2(k)$ is proportional to $1/c$. Furthermore, taking Taylor expansion in (\ref{BE2-r1}), we obtain
\begin{eqnarray}
r_1(k)&\approx &\frac{1}{2\pi}+\frac{1}{2\pi}\int_{-B_2}^{B_2}\frac{cr_2(k') }{\frac{c^2}{4}+k^2}\left[ 1-\frac{-2kk' +k'^2}{\frac{c^2}{4}+k^2}  \right] dk'\nonumber\\
&=&  \frac{1}{2\pi}\left[ 1+\frac{cn_{\downarrow}}{\frac{c^2}{4}+k^2}\right]+O(c^{-4}). \label{BE-Br1}
\end{eqnarray}
From the asymptotic distribution functions (\ref{BE-Br21}) and (\ref{BE-Br1}), we calculate the density 
\begin{eqnarray}
n&=&\int_{-B_1}^{B_1}r_1(k)  dk \approx \frac{B_1}{\pi}\left(1+\frac{4n_{\downarrow}}{c} -\frac{16B_1^2n_{\downarrow}}{3c^3}\right)\nonumber
\end{eqnarray}
that  gives 
\begin{equation}
B_1\approx n\pi\left(1-\frac{4n_{\downarrow}}{c}+\frac{16n_{\downarrow}^2}{c^2}+\frac{16n^2n_{\downarrow}\pi^2}{3c^3}-\frac{64n_{\downarrow}^3}{c^3}\right).\label{Q-r}
\end{equation}
From the energy (\ref{E2-r})  and the distribution function $r_1(k)$ (\ref{BE-Br1}),  we  may obtain an asymptotic   ground energy of the highly polarized  Fermi gas  with a strong repulsion ($c\gg B_1, \,B_2$)
\begin{eqnarray}
E &\approx &\frac{1}{3}n^3\pi^2\left[ 1-\frac{8n_{\downarrow}}{c} +\frac{48n_{\downarrow}^2}{c^2} \right.\nonumber\\
&&\left. -\frac{1}{c^3}\left(256n_{\downarrow}^3-\frac{32}{5}\pi^2n^2n_{\downarrow}\right)  \right]. \label{e-r-s}
\end{eqnarray}
In fact, for strong repulsion, the interacting energy in the ground state energy of the highly polarized Fermi gas solely depends  on the BA quantum number $N_{\downarrow}$,  A structure can be found  for 1D $\kappa$-component fermions \cite{note}.
By numerical checking,  we see that  for $\gamma>8$ and polarization $P=(N_{\uparrow}-N_{\downarrow})/(N_{\uparrow}+N_{\downarrow})\ge 0.5$, the energy (\ref{e-r-s}) is very  accurate,  see Fig. \ref{fig:Energy2}. 


\subsection{Weak  repulsion}

For weakly repulsive  regime,  it is convenient to rewrite the Fredholm equations (\ref{BE2-r1}) and (\ref{BE2-r2}) as 
\begin{eqnarray}
 {r_1}(k) &=& \frac{1}{2\pi}+\int_{-B_2}^{B_2}K_1(k-k'){r_2}(k') dk', \label{F2-F-r1}\\
 {r_2}(k) &= & \frac{1}{2\pi}-\int_{|k'|>B_1}K_1(k-k'){r_1} (k')dk'\label{F2-F-r2}
 \end{eqnarray}
 The derivation of (\ref{F2-F-r2}) is straightforward by using Fourier transform of the Fredholm equations (\ref{BE2-r1}) and (\ref{BE2-r2}), where for our convenience in the study, we 
 actually used
  \begin{eqnarray}
 {{r_m}}_{\rm in}(k)&=&\left\{  \begin{array}{ll}{r_m}(k)& |k|\le B_m\\
 0&|k|>B_m\end{array}\right. ,\nonumber\\
  {r_m}_{\rm out}(k)&=&\left\{  \begin{array}{ll}{r_m}(k)& |k|>B_m\\
 0&|k|\le B_m\end{array}\right. \label{R-in-out}
 \end{eqnarray}
with $m=1,2$ in Fourier transformation. These Fredholm equations are valid for arbitrary polarization including the balanced case. 
In the following unification of the ground state energy, we assume $B_1>B_2$ as an ansatz.   In the light of Takahashi's unification of the ground state energy \cite{Takahashi2},  we  give  the ground state energy per length
\begin{eqnarray}
E&=&\frac{B_1^2}{3\pi}+\frac{1}{2\pi}\int_{-B_2}^{B_2}H(k,B_1)dk \label{E-r1}\\
&&-\int_{-B_2}^{B_2}\left[ \int_{|k'|>B_1}K_1(k-k'){r_1}(k') dk'\right]H(k,B_1)dk,\nonumber
\end{eqnarray}
where 
\begin{eqnarray}
H(x,y)&=&\frac{1}{\pi}\left[(x^2-\frac{c^2}{4})\pi g_{y}(x)+yc \right.\nonumber\\
&&\left.+\frac{1}{2}x c \ln \frac{4(x-y)^2+c^2}{4(x+y)^2+c^2}\right],\nonumber\\
g_y(x)&=&1-G_+(y,x), \nonumber\\
G_{\pm}(x,y)&=&\tan^{-1}\frac{c}{2(x-y)}\pm \tan^{-1}\frac{c}{2(x+y)}. \nonumber
\end{eqnarray}
From  (\ref{repulsive-d}), we find that  the integration boundaries $B_1$ and $B_2$  satisfy the following conditions
\begin{eqnarray}
 \frac{N_{\uparrow}}{L}&=&\frac{B_1}{\pi}- \frac{1}{\pi}\int_{-B_2}^{B_2} r_2(k)  G_+(B_1,k)dk, \label{F2-r-n1}\\
\frac{N_{\downarrow} }{L} &=&  \frac{B_2}{\pi}- \frac{1}{\pi}\int_{|k|>B_1}r_1(k)  G_-(k,B_2) dk, \label{F2-r-n2}
 \end{eqnarray}
 in this weakly repulsive  regime.  Using  the condition (\ref{F2-r-n2}),  we may obtain   the  integration boundary $B_2$ (up to an order of  $c$ contributions),  
\begin{eqnarray}
B_2&\approx & n_{\downarrow}\pi + \frac{c}{4\pi}\ln\frac{4(B_1+B_2)^2+c^2}{4(B_1-B_2)^2+c^2}\nonumber\\
&& -\frac{ (B_1-B_2)}{\pi} \tan^{-1}\frac{c}{2(B_1-B_2)}\nonumber\\
&&+\frac{(B_1+B_2)}{\pi}\tan^{-1}\frac{c}{2(B_1+B_2)},\label{Q-r2}
\end{eqnarray}
or 
\begin{eqnarray}
B_2 &=& n_{\downarrow} \pi +\frac{c}{2\pi}\ln\frac{|B_1+B_2|}{|B_1-B_2|}+O(c^2).\label{B2}
\end{eqnarray}
The logarithm term in (\ref{Q-r2}) is converged as $B_1=B_2$. However,  the logarithm term  in (\ref{B2})  becomes  divergent  as $B_1=B_2$. This divergent  term in the ground state energy can be canceled out. Here we see a subtlety of this asymptotic expansion. 
Similarly, we calculate the Fermi momentum $B_1$ by definition (\ref{repulsive-d}) and the distribution (\ref{F2-F-r1})
\begin{eqnarray}
B_1&=& n_{\uparrow}\pi + \frac{c}{4\pi}\ln\frac{4(B_1+B_2)^2+c^2}{4(B_1-B_2)^2+c^2}\nonumber\\
&& -\frac{ (B_1-B_2)}{\pi} \tan^{-1}\frac{c}{2(B_1-B_2)}\nonumber\\
&&+\frac{(B_1+B_2)}{\pi}\tan^{-1}\frac{c}{2(B_1+B_2)},\label{Q-r1}
\end{eqnarray}
or
\begin{eqnarray}
B_1\approx n_{\uparrow}\pi +\frac{c}{2\pi}\ln\frac{|B_1+B_2|}{|B_1-B_2|}+O(c^2).\label{B1}
\end{eqnarray}
The ground state energy per length   in weakly repulsive  coupling limit can be expressed in terms of the  Fermi boundaries 
\begin{eqnarray}
E&\approx &\frac{B_1^3}{3\pi}+\frac{B_2^3}{3\pi}+\frac{2c}{\pi^2}B_1B_2\nonumber\\
&&-\frac{c}{2\pi^2}\left(B_1^2+B_2^2\right)\log\frac{|B_1+B_2|}{|B_1-B_2|}\label{E-bb}
\end{eqnarray}
Substituting (\ref{B2}) and  (\ref{B1}) into (\ref{E-bb}), we obtain the ground state energy of the Fermi gas with  a  weakly repulsive  interaction and with polarization  
\begin{eqnarray}
E=\frac{1}{3}n_{\uparrow}^3\pi^2+\frac{1}{3}n_{\downarrow}^3\pi^2+2cn_{\uparrow}n_{\downarrow}+O(c^2).\label{e-r-wub}
\end{eqnarray}
This leading order correction to the  interaction energy indicates a mean field effect.  By numerical checking,  we see that  the  energy  (\ref{e-r-wub}) agrees well with numerical result  in this weak coupling regime,  see Fig. \ref{fig:Energy2}.

 For balanced case,  the integration boundary $B_2$ tends to infinity.  It is different from above setting where we consider $B_1>B_2$.   Fredholm equations (\ref{BE2-r1}) and (\ref{BE2-r2}) (or from (\ref{F2-F-r1}) and (\ref{F2-F-r2})),   can be simplified with the help of Fourier transformation 
\begin{eqnarray}
r_1(k)&=&\frac{1}{\pi}-\int_{|k'|>B_1} K_2(k-k') {r_1}_{\rm out}(k') dk' .\label{r11}
\end{eqnarray}
However, for  $|k|>B_1$,  we find 
\begin{eqnarray}
{r_1}_{\rm out}(k)&=&\frac{1}{2\pi}-\int_{-\infty}^{\infty}R(k-k'){r_1}_{\rm in}(k') dk',\label{r1-out}
\end{eqnarray}
where the function $R(k)$ is given by 
\begin{eqnarray}
R(k)&=&\frac{1}{2\pi} \int_{-\infty}^{\infty}\frac{1}{1+e^{-|\omega|c}}e^{-\mathrm{i} \omega k}d\omega\nonumber\\
&=&-\frac{1}{\pi}\left( \frac{c}{4k^2}+\frac{c^3}{8k^4}+\cdots \right).\nonumber
\end{eqnarray}
We see clearly that the second term in (\ref{r1-out})  gives a contribution $O(c)$. Substituting the leading term  $r_1(k)=1/2\pi$ for the region $|k|>B_1$  into the distribution function $r_1(k)$ (\ref{r11}), we obtain the distribution function $r_1(k)$ for $|k|<B_1 $
\begin{eqnarray}
{r_1}(k) \approx \frac{1}{\pi}-\frac{1}{2\pi^2}\left[ \tan^{-1} \frac{c}{k+B_1} +\tan ^{-1} \frac{c}{B_1-k} \right].\nonumber 
\end{eqnarray}
From the relation $n=\int_{-B_1}^{B_1}{r_1}(k) dk$, we find 
\begin{eqnarray}
B_1&\approx &\frac{n\pi}{2}\left[1+\frac{c}{4\pi B_1}\log \frac{4B_1^2+c^2}{c^2}+\frac{1}{\pi}\tan^{-1} \frac{c}{2B_1} \right].\label{B-B1}
\end{eqnarray}
Then we obtain  the balanced ground state energy  for weakly repulsive interaction
\begin{eqnarray}
E&=&\frac{2}{3\pi}B_1^3\left[1-\frac{3}{4\pi}\left(\frac{c}{B_1} \log \frac{4B_1^2+c^2}{c^2}\right.\right. \nonumber \\
&&\left. \left.+\frac{4}{3}\tan^{-1}\frac{c}{2B_1}-\frac{8}{3}\frac{c}{B_1}\right)\right]+O(c^2). \label{E-b}
\end{eqnarray}
It is clearly seen that up to the order $O(c^2)$ the ground state energy of the balanced gas with a weak repulsion is converged  as $c\to 0$ (or say $cL/N\to 0$ by a rescaling in the above equations).  By  substituting $B_1$  into (\ref{E-b}),  we find  that the logarithmic  term   is  canceled .  The ground state energy for balanced case is given by 
\begin{equation}
E=\frac{1}{12}n^3\pi^2+\frac{1}{2}n^2c+O(c^2).\label{E-r-wb}
\end{equation}
We will further discuss the  continuity of the energy at a vanishing interaction strength next section.

\subsection{Weak  attraction}

For weakly attractive regime,  the Fredholm equations (\ref{Fermi2-a1}) and (\ref{Fermi2-a2}) are rewritten as
\begin{eqnarray}
 {\rho_1}(k) &=& \frac{1}{2\pi}+\int_{-Q_2}^{Q_2}K_1(k-k'){\rho_2}(k') dk', \label{F2-F-a1}\\
 {\rho_2}(k) &= & \frac{1}{2\pi}-\int_{|k'|>Q_1}K_1(k-k'){\rho_1} (k')dk'.\label{F2-F-a2}
\end{eqnarray}
These Fredholm equations are valid for arbitrary polarization.   It is seen clearly that  the  Fredholm equations  (\ref{F2-F-r1}) and    (\ref{F2-F-r2})   for repulsive regime and   the Fredholm equations (\ref{F2-F-a1}) and (\ref{F2-F-a2})   for attractive regime are identical as long as the integration boundaries match each other between the two sides. 
Similarly, for unification of the energy of the gas with a weak attraction, we assume $Q_1>Q_2$. In the above equations, $Q_1$ and $Q_2$ are determined by  
\begin{eqnarray}
 \frac{N_{\uparrow} }{L}&=&\frac{Q_1}{\pi}- \frac{1}{\pi}\int_{-Q_2}^{Q_2} \rho_2(k) G_+(Q_1,k)  dk,  \label{F2-a-n1}  \\
\frac{N_{\downarrow}}{L} &=&  \frac{Q_2}{\pi}-\frac{1}{\pi}\int_{|k|>Q_1} {\rho_1}(k)   G _-(k,Q_2) dk  \label{F2-a-n2}
\end{eqnarray}
that indeed match the integration boundaries $B_1$ and $B_2$   (\ref{F2-r-n1}), (\ref{F2-r-n2}) in the region $B_1>B_2$ and $Q_1>Q_2$.

Furthermore, substituting  (\ref{F2-F-a1}) and (\ref{F2-F-a2}) into the ground state energy  (\ref{Fermi2-E-a}), we obtain 
\begin{eqnarray}
E&=& \frac{Q_1^2}{3\pi}+\frac{1}{2\pi}\int_{-Q_2}^{Q_2}H(k,Q_1)dk\label{E-a1}\\
&&-\int_{-Q_2}^{Q_2}\left[ \int_{|k'|>Q_1}K_1(k-k'){\rho_1}(k')  dk'\right]H(k,Q_1)dk. \nonumber
\end{eqnarray}
From the equations (\ref{F2-a-n1}) and (\ref{F2-a-n2}), we find 
\begin{eqnarray}
Q_1&=& n_{\uparrow}\pi - \frac{|c|}{4\pi}\ln\frac{4(Q_1+Q_2)^2+c^2}{4(Q_1-Q_2)^2+c^2}\nonumber\\
&& +\frac{ (Q_1-Q_2)}{\pi} \tan^{-1}\frac{|c|}{2(Q_1-Q_2)}\nonumber\\
&&-\frac{(Q_1+Q_2)}{\pi}\tan^{-1}\frac{|c|}{2(Q_1+Q_2)}+O(c^2),\label{Q-a1}\\
Q_2&=& n_{\downarrow}\pi - \frac{|c|}{4\pi}\ln\frac{4(Q_1+Q_2)^2+c^2}{4(Q_1-Q_2)^2+c^2}\nonumber\\
&& +\frac{ (Q_1-Q_2)}{\pi} \tan^{-1}\frac{|c|}{2(Q_1-Q_2)}\nonumber\\
&&-\frac{(Q_1+Q_2)}{\pi}\tan^{-1}\frac{|c|}{2(Q_1+Q_2)}+O(c^2),\label{Q-a2}
\end{eqnarray}
Indeed, by calculating  the ground state energy (\ref{E-a1}) with the integration boundaries (\ref{Q-a1}) and (\ref{Q-a2})   for  weakly attractive interaction,  we do find a similar form of the ground state energy 
 \begin{eqnarray}
E=\frac{1}{3}n_{\uparrow}^3\pi^2+\frac{1}{3}n_{\downarrow}^3\pi^2-2|c|n_{\uparrow}n_{\downarrow}+O(c^2).\label{e-r-wbc}
\end{eqnarray}
 We see that the asymptotic ground state energy  (\ref{e-r-wub}) and  (\ref{e-r-wbc}) continuously connect at $c=0$ for arbitrary polarization, see Figures \ref{fig:Energy} and \ref{fig:Energy2}.    

 For balanced attractive regime, the Fermi boundaries $Q_1=0$ and $Q_2$ is finite,  the Fredholm equations (\ref{Fermi2-a1}) and (\ref{Fermi2-a2}) (or (\ref{F2-F-a1}) and (\ref{F2-F-a2})) reduce to 
 \begin{equation}
 \rho_2(k)=\frac{1}{\pi}+\int_{-Q_2}^{Q_2}K_2(k-k')\rho_2(k')dk'.
 \end{equation}
 By a iteration,  the Fermi boundary $Q_2$ is obtained from 
 $n=2\int_{-Q_2}^{Q_2}\rho_2(k)$ in a straightforward way
 \begin{eqnarray}
Q_2&\approx &\frac{n\pi}{2}\left[1-\frac{|c|}{4\pi Q_2}\log \frac{4Q_2^2+c^2}{c^2}-\frac{1}{\pi}\tan^{-1} \frac{|c|}{2Q_2}\right]\label{C-Q2}
\end{eqnarray}
that gives a similar form as  the Fermi boundary $B_1$ has, see  (\ref{B-B1}).  After a length algebra and iteration, we obtain the ground state energy 
\begin{eqnarray}
E&=&\frac{2}{3\pi}Q_2^3\left[1+\frac{3}{4\pi}\left(\frac{|c|}{Q_2} \log \frac{4Q_2^2+c^2}{c^2}\right.\right. \nonumber \\
&&\left. \left.+\frac{4}{3}\tan^{-1}\frac{|c|}{2Q_2}-\frac{8}{3}\frac{|c|}{Q_2}\right)\right]+O(c^2). \label{E-ca}
\end{eqnarray}
It is clearly seen that up to the oder $O(c^2)$ the ground state energy of the balanced gas with an attractive interaction is converged too as  $c\to 0^-$.  By  substituting $Q_2$  into (\ref{E-ca}),  we find  that the logarithmic term is  canceled out. Thus the energy is given by  
\begin{equation}
E=\frac{1}{12}n^3\pi^2-\frac{1}{2}n^2|c|+O(c^2)\label{E-a-wbb}
\end{equation}
that continuously connects to the energy (\ref{E-r-wb})  at $c\to 0$. But  the identity of the asymptotic expansions  may not mean that the energy analytically connects because of the divergence in the small region $c\to \mathrm{i} 0$ and the  mismatch of the Fermi boundaries associated the two sets of the Fredholm equations for both sides.  Nevertheless,   we see that under  a   mapping 
\begin{eqnarray}
r_1(k)\leftarrow \rightarrow \rho_1(k),\,\,\, r_2(k)\leftarrow \rightarrow \rho_2(k),\,\,\, c \leftarrow \rightarrow c,  \label{mapping2}
\end{eqnarray}
 the  Fredholm equations  (\ref{F2-F-r1}) and    (\ref{F2-F-r2}) with (\ref{F2-r-n1}), (\ref{F2-r-n2})  for repulsive regime and   the Fredholm equations (\ref{F2-F-a1}) and (\ref{F2-F-a2})  with   (\ref{F2-a-n1}), (\ref{F2-a-n2})  for attractive regime are identical for $Q_1>Q_2$ and $B_1>B_2$.   In the above equations $c>0$ for repulsive interaction  regime and $c<0$ for attractive interaction regime  are implied.  We also see that the ground state energy of the gas with a weakly repulsive interaction (\ref{E-r1}) and a weakly attractive interaction (\ref{E-a1})   are  unified under the mapping (\ref{mapping2}). This unification leads to the  continuity of the  energy  for this ploarized gas at vanishing interaction strength, i.e. $c\to 0$. Thus the asymptotic expansions of the energies of the repulsive and attractive Fermions  with non-zero polarization    are identical to all orders   in  the  vanishing interaction strength limits as long as the conditions $Q_1>Q_2$ and $B_1>B_2$ hold.  
 
 The analyticity  of the the energy at $c=0$ was discussed by Takahashi  \cite{Takahashi2}. Takahashi's theorem states that a) the energy function $f(n_{\uparrow}, n_{\downarrow};c)$ is analytic on the real axis of c when $n_{\uparrow}\ne n_{\downarrow}$; b)  $f(n_{\uparrow}, n_{\downarrow};c)$ is analytic on the real axis of c except for $c=0$ when $n_{\uparrow}= n_{\downarrow}$. This theorem appears not to be true for the region $B_1<B_2$ and $Q_1<Q_2$ in our study.  Takahashi's  proof of this theorem relies  on his  Lemma 2, i.e. the function $f$, density $n$ and density of spin-down Fermions $n_{\downarrow}$ are analytic as functions of $Q, B$ and $c$ except for the region $c=0$, and $Q<B$. Here $Q$ and $B$ are two integration boundaries.  Even the identity of the asymptotic expansions of the energy  may not mean the energy analytically connects  due to the divergence of the two sets Fredholm equations in the limit $c\to \mathrm{i}0$ and the mismatch of the intervals for the density distribution functions. Although we unified  the two sets of  Fredholm equations  (\ref{F2-F-r1}),    (\ref{F2-F-r2}) and (\ref{F2-F-a1}),  (\ref{F2-F-a2}) for arbitrary polarization,  the integration  boundaries  between the two regimes  are mismatched for the regions $Q_1<Q_2$ and $B_1<B_2$,  i.e.
 \begin{eqnarray}
 \frac{N_{\uparrow}}{L}&=&\frac{B_1}{\pi}- \frac{1}{\pi}\int_{-B_1}^{B_1} r_2(k)  G_+(B_1,k)dk\nonumber\\
 &&-\int_{B_1<|k|<B_2}r_2(k)\left[1-\frac{1}{\pi}G_-(k, B_1) \right]dk \label{F2-r-n1-2}\\
\frac{N_{\downarrow} }{L} &=&  \frac{B_2}{\pi}- \frac{1}{\pi}\int_{|k|>B_2}r_1(k)  G_-(k,B_2) dk \nonumber\\
&&-\int_{B_1<|k|<B_2}r_1(k)\left[1-\frac{1}{\pi}G_+(B_2,k) \right]dk\label{F2-r-n2-2}
 \end{eqnarray}
for weakly repulsive regime, and 
\begin{eqnarray}
 \frac{N_{\uparrow} }{L}&=&\frac{Q_1}{\pi}- \frac{1}{\pi}\int_{-Q_1}^{Q_1} \rho_2(k) G_+(Q_1,k)  dk\nonumber\\
 && +\int_{Q_1<|k|<Q_2}\rho_2(k) \left[1+\frac{1}{\pi}G_-(k,Q_1) \right] dk \label{F2-a-n1-2}  \\
\frac{N_{\downarrow}}{L} &=&  \frac{Q_2}{\pi}-\frac{1}{\pi}\int_{|k|>Q_2} {\rho_1}(k)   G _-(k,Q_2) dk\nonumber\\
&&+ \int_{Q_1<|k|<Q_2}\rho_1(k) \left[1+\frac{1}{\pi}G_+(Q_2,k)\right]dk\label{F2-a-n2-2}
\end{eqnarray}
 for weakly attractive regime. It is obvious the signs in the last term in each equation are mismatched.  In the above equations $c>0$ for repulsive interaction  regime and $c<0$ for attractive interaction regime  are implied.   This mismatch is clearly seen from the balanced case: $B_1 \to  Q_2$, $B_2 \to \infty$ and $ Q_1 \to 0$. Thus we see  that the Fredholm equations can not be  unified  for  the region $Q_1<Q_2$ and $B_1<B_2$ in the vanishing interaction strength.

\subsection{Strong  attraction}

In recent years,  strong attractive Fermi gas  has been received a considerable  attention  from theory and experiment  due to the existence of a novel pairing state. 
For the spin-1/2 Fermi gas with strongly attractive interaction, two Fermions with different spin states can form a tightly  bound pair. For the ground state, the model has three distinguished quantum phases, i.e. fully paired phase with equal number spin-up and -down Fermions, fully polarized phase of single spin-up Fermions and a partially polarized phase with both pairs and excess Fermions. The key features of this T = 0 phase diagram of the strongly attractive spin-1/2 Fermi gas were experimentally confirmed using finite temperature density profiles of trapped fermionic ${}^6$Li atoms \cite{Liao}.

Here we calculate the ground state energy (\ref{Fermi2-E-a}) from the   Fredholm equations (\ref{Fermi2-a1}) and (\ref{Fermi2-a2}) with the  integration boundaries $Q_1,Q_2$ that characterize the Fermi points of two Fermi seas, i.e. the Fermi seas for excess Fermions and pairs, respectively. Therefore,  for strong attraction, i.e. $|c|L/N\gg 1$, all integration boundaries are finite, i.e. $Q_1$ and $Q_2$ are finite.  In this regime, the conditions $c\gg Q_1,Q_2$  hold for arbitrary polarization. From the following calculation, we will see that the conditions $Q_1>Q_2$ and $Q_1<Q_2$ do not change the expression of the  energy.   Therefore,  the following result is   valid for arbitrary polarization, including the balance case.  In this regime, it is convenient to use a notation $|c|$ instead of a negative value of $c$.  The ground state energy is calculated in the following way
\begin{eqnarray}
E&=&\frac{Q^3_1}{3\pi} +\frac{1}{\pi}\int^{Q_2}_{-Q_2}\rho_2(k)\left[\left(k^2-\frac{c^2}{4}\right)\left(2\pi \right. \right. \nonumber \\
&&\left. \left. -\tan^{-1}\frac{2(Q_1-k)}{|c|}+  \tan^{-1}\frac{2(Q_1+k)}{|c|} \right) \right. \nonumber\\
&&\left. -Q_1|c|-\frac{1}{2}\lambda |c| \ln \frac{4(k-Q_1)^2+c^2}{4(k+Q_1)^2+c^2}\right]dk. \label{E-as}
\end{eqnarray}
Furthermore, we consider strong coupling expansion in  the energy (\ref{E-as}),  here we assume $|c|\gg Q_1, Q_2$. We collect contributions up to the order of $1/|c|^3$, i.e. 
\begin{eqnarray}
E&\approx &  \frac{Q_1^3}{3\pi}\left[1-\frac{4n_{\downarrow}}{|c|} +\frac{48Q_1^2n_{\downarrow}}{5|c|^3}+\frac{32Q_2^3}{3\pi|c|^3} \right]\nonumber\\
&&-\frac{c^2}{2}n_{\downarrow}+2\int_{-Q_2}^{Q_2} \rho_2(k)k^2dk. \label{E-aa}
\end{eqnarray}
In the last  equation of (\ref{E-aa}), the first part in the square  bracket  is  the kinetic energy of excess single atoms including marginal interference effect between the single atoms and molecules of two-atoms. The second term is the total binding energy of the bound pairs. The last term characterizes the total energy of the molecules of two-atom. We now  calculate the Fermi momenta $Q_1$ and $Q_2$ and the energy of the molecules of two-atom. For our convenience, we denote 
\begin{equation}
E=E_0^u+E_0^b+n_{\downarrow}\varepsilon_b
\end{equation}
 with
\begin{eqnarray}
E_0^u& =&\frac{Q_1^3}{3\pi}\left[1-\frac{4n_{\downarrow}}{|c|} +\frac{48Q_1^2n_{\downarrow}}{5|c|^3}+\frac{32Q_2^3}{3\pi|c|^3} \right],\nonumber\\
E_0^b&=& 2\int_{-Q_2}^{Q_2}\rho_2(k)k^2dk,\qquad \varepsilon_b =-\frac{c^2}{2}. \label{E0b}
\end{eqnarray}  

We calculate $Q_1$ from (\ref{density-a}) with  the density (\ref{Fermi2-a1}):
\begin{eqnarray}
n_{\uparrow}-n_{\downarrow}&=&\int_{-Q_1}^{Q_1}\left(\frac{1}{2\pi}-\int_{-Q_2}^{Q_2}K_1(k-k')\rho_2(k') \right)dk\nonumber\\
&\approx & \frac{Q_1}{\pi}\left[1-\frac{4n_{\downarrow}}{|c|}+ \frac{16Q_1^2n_{\downarrow}}{3|c|^3}  +\frac{32Q_2^3}{3\pi|c|^3} \right].\nonumber
\end{eqnarray}
Then we obtain  the Fermi momentum 
\begin{eqnarray}
Q_1&\approx& (n-n_{\downarrow})\pi \left[1+\frac{4n_{\downarrow}}{|c|}+\frac{16n_{\downarrow}^2}{c^2}\right.\nonumber\\
&&\left. -\frac{16}{3|c|^3}\left((n_{\uparrow}-n_{\downarrow})^2\pi^2n_{\downarrow}+\frac{n_{\downarrow}^3\pi^2}{4} -12n_{\downarrow}^3\right)  \right].\label{Qa}
\end{eqnarray}
Similarly, we calculate  $Q_2$ from (\ref{density-a})  with the distributions (\ref{Fermi2-a1}) and (\ref{Fermi2-a2})
\begin{eqnarray}
Q_2&\approx &\frac{n_{\downarrow }\pi}{2}\left(1+\frac{2n_{\uparrow}-n_{\downarrow}}{|c|}+ \frac{(2n_{\uparrow}-n_{\downarrow})^2}{c^2}+  \frac{(2n_{\uparrow}-n_{\downarrow})^3}{|c|^3}\right.\nonumber\\
&& \left. -\frac{n_{\downarrow}^2\pi^2(8n_{\uparrow}-7n_{\downarrow})}{12|c|^3}-\frac{\pi^2\left[n_{\downarrow}^3+32(n_{\uparrow}- n_{\downarrow})^3\right]}{12|c|^3}\right).\label{Ba}
\end{eqnarray}

We observe that  the kernels in the Fredholm equations (\ref{Fermi2-a1}) and (\ref{Fermi2-a2})  converge quickly with  the distribution functions as  the interacting strength $|c|$ increases. This allows one to take  a proper Taylor series expansion in the distribution functions. In this way,  from Eq. (\ref{Fermi2-a2}) we may  obtain  
\begin{eqnarray}
&&\rho_2 (k) \approx \frac{1}{\pi}\left[1-\frac{n_{\downarrow}|c|}{c^2+k^2}+ \frac{|c|E_0^b}{2(c^2+k^2)^2}  \right]\nonumber\\
&&-\frac{1}{2\pi}\left[\frac{|c|(n_{\uparrow}-n_{\downarrow})}{\frac{c^2}{4}+k^2}+\frac{|c|}{\left(\frac{c^2}{4}+k^2\right)^2} \int_{-Q_1}^{Q_1}\rho_1(k')k'^2dk'\right]\nonumber\\
&=&\frac{1}{\pi}+\frac{2Q_2^3}{3\pi^2|c|^3}+\frac{8Q_1^3}{3\pi^2|c|^3}-\frac{n_{\downarrow}}{\pi}\frac{|c|}{c^2+k^2}\nonumber\\
&&-\frac{n_{\uparrow}-n_{\downarrow}}{2\pi}\frac{|c|}{\frac{c^2}{4}+k^2}.\label{sigma-aa}
\end{eqnarray}
Substituting (\ref{sigma-aa}) into the energy $E_0^b$ (\ref{E0b}),
\begin{eqnarray}
E_0^b&\approx & \frac{4Q_2^3}{3\pi}+\frac{8Q_2^6}{9\pi^2|c|^3}+\frac{32Q_1^3Q_2^3}{9\pi^2|c|^3}\nonumber\\
&&-\frac{2n_{\downarrow}}{\pi}\int_{-Q_2}^{Q_2}\frac{|c|k^2dk}{c^2+k^2}
-\frac{(n_{\uparrow} -n_{\downarrow})}{\pi}\int_{-Q_2}^{Q_2}\frac{|c|k^2dk}{\frac{c^2}{4}+k^2}\nonumber\\
&\approx &\frac{4Q_2^3}{3\pi}\left(1-\frac{2n_{\uparrow}-n_{\downarrow}}{|c|}+\frac{2(Q_2^3+4Q_1^3)}{3\pi|c|^3}\right.\nonumber\\
&&\left.+\frac{3(8n_{\uparrow}-7n_{\downarrow})Q_2^2}{5|c|^3}\right).\label{E0-b}
\end{eqnarray}
Substituting equations (\ref{Qa}) and (\ref{Ba}) into the ground state energy (\ref{E-aa}) and (\ref{E0-b}), we obtain the ground state energy of the gas with a  strongly attractive  interaction and  with an arbitrary polarization
\begin{eqnarray}
E_0^u& \approx &\frac{(n_{\uparrow}-n_{\downarrow})^3\pi^2}{3}\left[1+\frac{8n_{\downarrow}}{|c|}+\frac{48n_{\downarrow}^2}{c^2}\right.\nonumber\\
&&\left. -\frac{8n_{\downarrow}}{15|c|^3}\left(12\pi^2(n_{\uparrow}-n_{\downarrow})^2-480 n_{\downarrow}^2+5n_{\downarrow}^2\pi^2\right) \right],\label{E0u-sa}\\
E_0^b& \approx& \frac{n_{\downarrow}^3\pi^2}{6}\left[1+\frac{2(2n_{\uparrow}-n_{\downarrow})}{|c|}  +\frac{3(2n_{\uparrow}-n_{\downarrow})^2}{c^2} \right.\nonumber\\
&&\left.  -\frac{4}{15|c|^3} \left(
180n_{\downarrow}n_{\uparrow}^2+20\pi^2n_{\uparrow}^3-90n_{\uparrow}n_{\downarrow}^2-22\pi^2n_{\downarrow}^3\right.\right.\nonumber\\
&&\left.\left.+15n_{\downarrow}^3-120n_{\uparrow}^3+63\pi^2n_{\downarrow}^2n_{\uparrow}-60\pi^2n_{\downarrow}n_{\uparrow}^2\right) \right].\label{E0b-sa}
\end{eqnarray}
We define polarization $P=(N_{\uparrow}-N_{\downarrow})/N=(n_{\uparrow}-n_{\downarrow})/n$,  then   the energy in terms of polarization is given by 
\begin{eqnarray}
E&\approx&\frac{\hbar^2n^3}{2m}\left\{-\frac{(1-P)\gamma^2}{4}
+\frac{\pi^2(1-3P+3P^2+15P^3)}{48}\right.\nonumber\\
&&\left.+\frac{\pi^2(1-P)(1+P-5P^2+67P^3)}{48|\gamma|}\right. \nonumber\\
&&\left. +\frac{\pi^2(1-P)^2(1+5P+3P^2+247P^3)}{64\gamma^2}\right.\nonumber\\
&&\left.-\frac{\pi^2(1-P)}{1440|\gamma|^3}\left[-15+31125{P}^{4}+1861{\pi }^{2}{P}^{5}\right.\right.\nonumber\\
&&\left.\left. -15765P^5-659{\pi }^{2}{P}^{4} +346{\pi }^{2}{P}^{3}-14{\pi }^{2}{P}^{2} \right.\right.\nonumber\\
& &\left.\left.+\pi^2
P+{\pi }^{2}-105P-150P^2-15090P^3\right]\right\}.\label{Energy}
\end{eqnarray}
that  agrees  with the result derived from dressed energy equations \cite{Guan,Wadati}.
This result is highly accurate  as being  seen in Fig. \ref{fig:Energy} and Fig. \ref{fig:Energy2}.
From the energies (\ref{E0u-sa}) and (\ref{E0b-sa}), we see the bound pairs have tails and the interfere with each other. But, it is impossible to separate the intermolecular forces from the interference between  molecules and  single Fermions. If we consider 
$n_{\downarrow}\gg x=n_{\uparrow}-n_{\downarrow}$,  the single atoms are repelled by the molecules, i.e., 
\begin{eqnarray}
E(n_{\downarrow}, x)\approx \frac{E(n_{\downarrow},0)}{L}+\frac{1}{6}n_{\downarrow }^3\pi^2\left[\frac{4x}{|c|} +\frac{12x(x+n_{\downarrow})}{c^2}\right].\nonumber 
\end{eqnarray}
Where 
\begin{eqnarray}
E(n_{\downarrow}, 0)\approx  \frac{1}{6}n_{\downarrow }^3\pi^2\left(1+\frac{2n_{\downarrow}}{|c|}+\frac{3n_{\downarrow}^2}{c^2}\right)+\varepsilon_b.\nonumber 
\end{eqnarray}

In addition, the phase diagram and magnetism can be work out directly from the relations (\ref{chemical}) with the ground state energy for the four regimes. The phase boundaries of the  full  phase diagrams may be analytically and numerically  obtained by imposing the conditions $s_z=0,\, 0.5$ in the conditions (\ref{chemical}), which has been discussed in literature \cite{Orso,HuiHu,Guan,colome}.  

\section{Conclusion}

In conclusion, we have presented  a systematic method to   derived the first few of terms of the asymptotic expansion of  the Fredholm equations for the spin-1/2 Fermi gas with repulsive delta-function and attractive delta-function  interactions in  the four regimes:   A) strongly repulsive regime; B) weakly repulsive regime; C) weakly attractive regime and D) strongly attractive regime.   We have obtained explicitly the  ground state energy  of the Fermi gas with  polarization  in these regimes, see the key result (\ref{E-A-b}), (\ref{e-r-s}), (\ref{e-r-wub}),  (\ref{e-r-wbc}), (\ref{E0u-sa}) and (\ref{E0b-sa}).
By numerical checking, these asymptotic  ground state energy are  seen to be highly accurate in the four regimes. In weakly attractive and repulsive regimes, the ground state energies, integration boundary relations and the associated  two sets of the Fredholm equations have been unified.  The  two sets of the Fredholm equations can be identical as long as the associated integration boundaries match each other between  the two sides. This suggests that  the asymptotic expansions of the energies of the repulsive and attractive Fermions   are identical to all orders  in this region as $c\to 0$.  The identity of the asymptotic expansions  may not mean that the energy analytically connects due to the divergence of the two sets Fredholm equations in the limit $c\to \mathrm{i}0$ and the mismatch of the associated 
integration boundaries between two sides at some intervals, e.g.,  $B_{1}<B_{2}$ and $Q_{1}<Q_{2}$.   

Moreover, the explicit result of   the ground state energy obtained provides facilities to study  universal nature of many-body phenomena.  The local pair correlation for opposite spins  can be calculated directly from   the ground state energy  by  
\begin{equation} 
g^{(2)}_{\uparrow,\downarrow}(0) =\frac{1}{2n_{\uparrow}n_{\downarrow}}\partial E(n,s_z)/\partial c.\nonumber 
\end{equation}
  This naturally gives the 1D analog of the Tan's adiabatic theorem \cite{Tan}  through the relation  
  \begin{equation}
  {\cal C}=\frac{4}{a_{1D}^2}n_{\uparrow}n_{\downarrow}g^{(2)}_{\uparrow,\downarrow}(0),\nonumber 
  \end{equation}
 where ${\cal C}$ is called the universal contact, measuring the probability that two fermions with opposite spin stay together.  It was shown \cite{Tan}  that the momentum distribution exhibits universal ${\cal C}/k^4$ decay as the momentum tends to infinity.
The significant feature of Tan's universal contact is that it can be applied to any many-body system of interacting bosons and fermions in 1D, 2D and 3D  \cite{Tan,Zwerger2}.  In addition, the explicit forms of the ground state energies in the four regimes can  be used to determine magnetism and phase diagram of the system in the grand canonical ensemble.  It can help conceive quantum statistical effect by a comparison between the ground state energies of 1D delta-function interacting  fermions and spinless bosons.   These provide a precise understanding of many-body correlations and quantum magnetism in the context of cold atoms.  The method which we have developed in this paper can be generalized to study ground state properties of  1D multicomponent Fermi and Bose gases with delta-function interaction. We will consider it  in  \cite{note}.

{\bf Acknowledgment.}  The authors thank Professor Chen Ning Yang for initiating this topic and for valuable  discussions and suggestions. They also thank Dr. Xiang-Guo Yin for doing numerical checking and preparing the figures.   The author XWG  has been supported by the Australian Research Council. This work is partly  supported  by  Natural Science Foundation of China under the grants No. 11075014 and No. 11174099.


\end{document}